\documentclass[12pt,twoside]{article}
\usepackage[english]{babel}
\usepackage{latexsym}
\usepackage{amsmath}
\usepackage{amsfonts}
\usepackage{graphicx}
\usepackage{natbib}

\newtheorem{theorem}{Theorem}[section]
\newtheorem{lemma}[theorem]{Lemma}
\newtheorem{corollary}[theorem]{Corollary}
\newtheorem{definition}[theorem]{Definition}
\newtheorem{proposition}[theorem]{Proposition}

\newtheorem{example}[theorem]{Example}

\newcommand{\qed}{\hfill\ensuremath{\Box}}
\newenvironment{proof}[1][Proof]{\begin{trivlist}\item[\hskip \labelsep {\bfseries #1}]}{\qed\end{trivlist}}

\newcommand{\pdgmaxent}{\ensuremath{\pdgentropy_{q}(\tilde{p})}}

\newcommand{\pdgnulvec}{\mathbf{0}} 
\newcommand{\pdgssrv}{\ensuremath{T}}
\newcommand{\pdgowninf}[2]{\ensuremath{\inf \; \{#1 \mid #2 \}}}
\newcommand{\pdgnatpar}{\ensuremath{\tilde{\beta}}}
\newcommand{\pdgprobset}{\ensuremath {{\mathcal P}}}
\newcommand{\pdgnlim}{\ensuremath{\lim_{n \rightarrow \infty}}}
\newcommand{\pdgaverage}[2]{\ensuremath{\overline{{#1}^{#2}}}}
\newcommand{\pdgindicator}{\ensuremath{I}}
\newcommand{\pdgxsamplespace}{\ensuremath{{\cal X}}}
\newcommand{\pdgmyconstraint}{\ensuremath{\pdgmyavg = \pdgssval}}
\newcommand{\pdgset}[1]{\ensuremath{{\cal #1}}}
\newcommand{\pdgstsum}{\sum_{i=1}^n}

\newcommand{\pdgmyavg}{\ensuremath{\pdgaverage{\pdgssfn}{(n)}}}
\newcommand{\pdgssfn}{\ensuremath{T}}
\newcommand{\pdgnormde}{\ensuremath{\aleph}} 

\newcommand{\pdgoutcome}{x}
\newcommand{\pdgentropy}{{\ensuremath{\mathbf H}}}
\newcommand{\pdgExp}[2]{E_{#1} [#2]}
\newcommand{\pdgreals}{{\bf R}}
\newcommand{\pdgnaturals}{{\bf N}}

\newcommand{\pdgssval}{\ensuremath{\tilde{t}}}

\begin{document}
\begin{titlepage}
  \title{Entropy Concentration and the Empirical Coding Game}
  \author{\Large{Peter Gr\"unwald}\thanks{Also: research fellow at
      EURANDOM, P.O. Box 513, 5600 MB Eindhoven, The Netherlands. This
      is a slightly modified version of the paper with the same title
      that appeared in {\em Statistica Neerlandica\/} 62(3), 2008,
      pages 374--392,  on the occasion
      of the 10th anniversary of EURANDOM. Some of the results presented
here have already appeared in the conference paper
\citep{Grunwald01a} and the technical report \citep{Grunwald01b}. Theorems~\ref{thm:A} and~\ref{thm:B} of Section~\ref{sec:applications} are new and
have not been published before. The paper benefited enormously from various
      discussions with Richard Gill, Phil Dawid and Franz Merkl. This
      work was supported in part by the IST Programme of the European
      Community, under the PASCAL Network of Excellence,
      IST-2002-506778. This publication only reflects the author's
      views.}\\[5mm]
    {\em CWI, P.O. Box 94079,1090 GB Amsterdam, NL}
}
\date{}
\end{titlepage}
\maketitle
\begin{abstract}
We give a characterization of Maximum Entropy/Minimum Relative
  Entropy inference by providing two `strong entropy concentration'
  theorems. These theorems unify and generalize Jaynes'
  `concentration phenomenon' and Van Campenhout and Cover's
  `conditional limit theorem'. The theorems characterize exactly in
  what sense a prior distribution $Q$ conditioned on a given
  constraint and the distribution $\tilde{P}$ minimizing $D(P || Q)$ over
  all $P$ satisfying the constraint are `close' to each other. We then apply our theorems to establish
  the relationship between entropy concentration and a
  game-theoretic characterization of Maximum Entropy Inference due to
  Tops{\o}e and others.
\end{abstract}

\section{Introduction}
Jaynes' Maximum Entropy (MaxEnt) Principle is a well-known principle for inductive
inference
\citep{Csiszar75,Csiszar91,Topsoe79,CampenhoutC81,CoverT91,GrunwaldD04}.
It has been applied to statistical and machine learning
problems ranging from protein modeling to stock market prediction \citep{KapurK92}. One of
its characterizations (some would say `justifications') is the
so-called {\em concentration phenomenon\/}
\citep{Jaynes78,Jaynes82}. Here is an informal version of this phenomenon, 
in the words of \cite{Jaynes03}:
\begin{quote}
``If the information incorporated into the maximum-entropy analysis
includes all the constraints actually operating in the random
experiment, then the distribution predicted by maximum entropy is
overwhelmingly the most likely to be observed experimentally.''
\end{quote}
For the case in which a prior distribution over the domain at hand is
available, \cite{CampenhoutC81} have
proven the related {\em conditional limit theorem}.
In Sections~\ref{sec:overview}-\ref{sec:concentration}, we
provide a strong generalization
of both the concentration phenomenon and the conditional limit
theorem. In Section~\ref{sec:applications}, the results of
Section~\ref{sec:concentration} are used to  
extend an existing game-theoretic characterization (again, some would
say ``justification'')  of Maximum Entropy
due to \cite{Topsoe79}. 
In this way, we provide sharper results on two of the most frequently 
cited characterizations of the maximum entropy principle.
\section{Informal Overview}
\label{sec:overview}
\paragraph{Maximum Entropy}
Let $X$ be a random variable taking values in some set
$\pdgxsamplespace$, which (only for the time being!) we assume to be finite:
$\pdgxsamplespace = \{1, \ldots, m \}$. 
Let $P, Q$ be distributions for $\pdgxsamplespace$ with probability mass
functions $p$
and $q$. We define
$\pdgentropy_{Q}(P)$, {\em the $Q$-entropy
of $P$}, as 
\begin{equation}
\label{eq:maxentdef}
\pdgentropy_{Q}(P) = - E_{P} \left[\log \frac{p(x)}{q(x)}\right] = - D(P || Q),
\end{equation} 
where $D(\cdot \| \cdot)$ is the Kullback-Leibler (KL) divergence
between $P$ and $Q$ \citep{CoverT91}. 
In the usual MaxEnt setting, we are
given a `prior' distribution $Q$ and a {\em moment constraint\/}:
\begin{equation}
\label{eq-constraint}
\pdgExp{ }{T(X)} = \tilde{t}
\end{equation}
where $T$ is some function $T: \pdgxsamplespace \rightarrow
\pdgreals^k$ for some $k>0$ (More general formulations with
  arbitrary convex constraints exist \citep{Csiszar75}, but here we
  stick to constraints of form (\ref{eq-constraint})).
We define, if it exists, $\tilde{P}$ to be the unique distribution over
$\pdgxsamplespace$ that maximizes the $Q$-entropy 
over all
distributions (over $\pdgxsamplespace$) satisfying (\ref{eq-constraint}):
\begin{equation}
\label{eq:argmax}
\tilde{P} =  \arg \max_{\{ P : E_P[T(X)] = \tilde{t}\}} 
\pdgentropy_{Q}(P) = \arg \min_{\{ P : E_P[T(X)] = \tilde{t}\}} 
D(P || Q)
\end{equation}
The MaxEnt Principle then tells us that, in absence of any
further knowledge about the `true' or `posterior' distribution
according to which
data are distributed, our best guess for it is $\tilde{P}$.
In practical problems we are usually not given a constraint of form
(\ref{eq-constraint}). Rather we are given 
an {\em empirical constraint\/} of
the form
\begin{equation}
\label{eq-emp}
\frac{1}{n} \pdgstsum T(X_i)  = \tilde{t} \mbox{\ \ \ which we always abbreviate to\ `$
\pdgmyavg = \tilde{t}$'}
\end{equation}
The MaxEnt Principle is then usually applied as follows: suppose we are
given an empirical constraint of form (\ref{eq-emp}). We then have to
make predictions about new data coming from the same source. In
absence of knowledge of any `true' distribution generating this data,
we should make our predictions based on the MaxEnt distribution
$\tilde{P}$ for the moment constraint (\ref{eq-constraint}) corresponding to
empirical constraint (\ref{eq-emp}). $\tilde{P}$ is extended to several outcomes by taking the
product distribution. 
\paragraph{The Concentration Phenomenon and The Conditional Limit
  Theorem} Why should this procedure make any sense? Here
is one justification.
If $\pdgxsamplespace$ is finite, and in the absence of any prior
knowledge beside the constraint, 
one usually picks the uniform distribution for $Q$. In this case, Jaynes'
`concentration phenomenon' applies (We are referring here to
  the version employed by \cite{Jaynes78}. The theorem of \cite{Jaynes82}
  extends this in a direction different from the one we consider here). It says that 
for all
$\varepsilon > 0$, 
\begin{equation}
\label{eq:snack}
Q^n \left(\sup_{j \in \pdgxsamplespace} \left| \frac{1}{n} \pdgstsum
\pdgindicator_{j}(X_i) - \tilde{P}(X=j) \right| > \varepsilon \ \mid \ \pdgmyavg =
\tilde{t} \right) = O(e^{-cn})
\end{equation}
for some constant $c$ depending on $\varepsilon$.
Here $Q^n$ is the $n$-fold product 
distribution of $Q$, and $\pdgindicator$ is the indicator function: $\pdgindicator_j(x) = 1$ if
$x= j$ and $0$ otherwise. In words, for the
overwhelming majority among the sequences satisfying the constraint,
the empirical frequencies are close to the maximum entropy
probabilities. It turns out that (\ref{eq:snack}) still holds if $Q$ is
non-uniform. For an illustration we refer to Example~\ref{ex:brandeis}.  
A closely related result (Theorem 1 of  \cite{CampenhoutC81}) is the
conditional limit theorem (This theorem too
  has later been extended in several directions different from the one
  considered here; see the discussion at the end of 
Section~\ref{sec:concentration}). It says that
\begin{equation}
  \label{eq:campenhout}
  \lim_{\stackrel{\scriptstyle n \rightarrow \infty}{n \tilde{t} \in \pdgnaturals}}
Q^1(\cdot \mid \pdgmyconstraint) = \tilde{P}^1(\cdot)
\end{equation}
where $Q^1(\cdot \mid \pdgmyconstraint)$ and $\tilde{P}^1(\cdot)$ refer
to the marginal distribution of $X_1$ under $Q(\cdot \mid \pdgmyconstraint)$ and $\tilde{P}$  respectively.
\paragraph{Our Results}
Both theorems above say that 
for some sets $\pdgset{A}$,
\begin{equation}
  \label{eq:hertz}
  Q^n(\pdgset{A} \mid \pdgmyconstraint) \approx \tilde{P}^n(\pdgset{A})
\end{equation}
In the concentration phenomenon, the set $\pdgset{A} \subset
\pdgxsamplespace^n$ is about the frequencies of individual outcomes in
the sample. In the conditional limit theorem $\pdgset{A} \subset
\pdgxsamplespace^1$ only concerns the first outcome.  One might
conjecture that (\ref{eq:hertz}) holds asymptotically in a much wider
sense, namely for {\em just about any set whose probability one may be
  interested in}. For examples of such sets see
Example~\ref{ex:brandeis}.  In Theorems~\ref{th-discrete}
and~\ref{th-general} we show that (\ref{eq:hertz}) indeed holds for a
very large class of sets; moreover, we give an explicit indication of
the error one makes if one approximates $Q(\pdgset{A} \mid
\pdgmyconstraint)$ by $\tilde{P}(\pdgset{A})$. In this way we unify
and strengthen both the concentration phenomenon and the conditional
limit theorem. To be more precise, let $\{\pdgset{A}_n\}$, with
$\pdgset{A}_i \subset \pdgxsamplespace^i$ be a sequence of `typical'
sets for $\tilde{P}$ in the sense that $\tilde{P}^n(\pdgset{A}_n)$
goes to 1 sufficiently fast. Then broadly speaking
Theorem~\ref{th-discrete} shows that $Q^n(\pdgset{A}_n \mid \pdgmyavg
= \tilde{t})$ goes to 1 too, `almost' as fast as
$\tilde{P}^n(\pdgset{A}_n)$.  Theorem~\ref{th-general}, our main
theorem, says that, if $m_n$ is an arbitrary increasing sequence with
$\pdgnlim m_n/n = 0$, then for {\em every\/} (measurable) sequence
$\pdgset{A}_{m_1}, \pdgset{A}_{m_2}, \ldots$ (i.e. not just the
typical ones), with $\pdgset{A}_{m_n} \subset \pdgxsamplespace^{m_n}$,
$\tilde{P}^n(\pdgset{A}_{m_n} ) \rightarrow Q^n(\pdgset{A}_{m_n} \mid
\pdgmyconstraint)$.  In Section~\ref{sec:applications}, we first give
an interpretation of our strong concentration results in terms of data
compression. We then show (Theorem~\ref{th:infsup}) that our
concentration phenomenon implies that the MaxEnt distribution
$\tilde{P}$ achieves the best minimax time-averaged logarithmic loss
(codelength) achievable for sequential prediction of samples
satisfying the constraint. We also characterize (Theorem~\ref{thm:A}
and~\ref{thm:B}) the precise conditions under which $\tilde{P}$ also
achieves the total (non-time averaged) minimax logarithmic loss.
Surprisingly, the answer depends crucially on the dimensionality $k$
of the constraint random vector $T$: for $k \leq 2$, $\tilde{P}$ is
also best in the total sense. For $k > 3$, there exist distributions
which consistently outperform $\tilde{P}$. This is related to the
well-known fact that random walks in $\pdgreals^k$ are transient if $k
\geq 3$.
\section{Mathematical Preliminaries}
\paragraph{The Sample Space}
From now on we  assume a sample space $\pdgxsamplespace \subseteq \pdgreals^l$ for some
$l > 0$  and let $X$ be the random vector with $X(x) = x$ for all $x
\in \pdgxsamplespace$. We reserve the symbol $Q$ to refer to a
distribution for $X$ called the {\em prior distribution\/} (formally,
$Q$ is a distribution on $(\pdgxsamplespace,\sigma(X))$ where
$\sigma(X)$ is the Borel-$\sigma$-algebra generated by $X$).
We will be interested in sequences of i.i.d. random variables $X_1,
X_2, \ldots$,
all distributed according to $Q$. Whenever no confusion can arise,
we use $Q$ also to refer to the joint
(product) distribution of $\times_{i \in \pdgnaturals} X_i$. Otherwise, we use
$Q^m$ to denote the $m$-fold product distribution of $Q$. The sample $(X_1, \ldots, X_m)$ will also be written as
$X^{(m)}$.
\paragraph{The Constraint Functions $T$}
Let $T = (T_{[1]}, \ldots, T_{[k]})$ be a $k$-dimensional random vector that is
$\sigma(X)$-measurable. We refer to the event $\{ x \in \pdgxsamplespace
\mid T(x) =t\}$ both as `$T(X) = t$' and as `$T = t$'.
Similarly we write $T_i = t$ as an abbreviation of $T(X_i) = t$ and
$T^{(n)}$ as short for $(T(X_1), \ldots, T(X_n))$.
The {\em average\/} of $n$ observations of $T$ will be denoted by
$\pdgaverage{T}{(n)} := {n}^{-1} \pdgstsum  T(X_i).$
We assume that the support of 
$\pdgxsamplespace$ is either countable (in which case the prior
distribution $Q$ admits a probability mass function) or that it is
a connected subset of $\pdgreals^l$ for some $l > 1$ (in which case we assume that $Q$ has
a bounded continuous density with respect to Lebesgue measure). In both cases, we denote
the probability mass function/density  by  $q$. If $\pdgxsamplespace$
is countable, we shall further assume that $T$ is of the {\em lattice
  form\/} (which it will be in most applications):
\begin{definition}{\bf \cite[Page 490]{Feller68b}}
\label{def:lattice}
A {\em $k$-dimensional lattice random vector\/} \\
$T = (T_{[1]}, \ldots,
T_{[k]})$ is a random vector for which there exists real-valued $b_1,
\ldots, b_k$ and $h_1, \ldots,
h_k$ such that, for $1 \leq j \leq k$, 
$\forall x \in \pdgxsamplespace:$
$T_{[j]}(x) \in \{ b_j + s h_j\ \mid s \in \pdgnaturals \}$.
We call the largest $h_i$ for which this holds the {\em
span\/} of $T_{[i]}$.
\end{definition}
If $X$ is continuous, we shall assume that $T$ is `regular':
\begin{definition}
We say a  $k$-dimensional random vector  is of {\em regular continuous
  form\/} if its distribution  
admits a bounded continuous density with respect to Lebesgue measure.
\end{definition}
\paragraph{Maximum Entropy}
Throughout the paper, $\log$ is used to denote logarithm to base
2. Let $P, Q$ be distributions for $\pdgxsamplespace$. We define
$\pdgentropy_{Q}(P)$, {\em the $Q$-entropy
of $P$}, as 
\begin{equation}
\pdgentropy_{Q}(P) = - D(P || Q),
\end{equation} 
where $D$ is the KL-divergence between $P$ and $Q$.
This is defined even if $P$ or $Q$ have no densities 
\citep{Csiszar75}.  Assume we are given a constraint of form
(\ref{eq-constraint}), i.e.  $\pdgExp{P}{T(X)} = \tilde{t}$.  Here $T
= (T_{[1]},\ldots, T_{[k]}), \tilde{t} = (\tilde{t}_{[1]}, \ldots,
\tilde{t}_{[k]})$.  We define, if it exists, $\tilde{P}$ to be the
unique distribution on $\pdgxsamplespace$ that maximizes the
$Q$-entropy over all distributions on $\pdgxsamplespace$ satisfying
(\ref{eq-constraint}). That is, $\tilde{P}$ is given by
(\ref{eq:argmax}).  If Condition 1 below holds, then $\tilde{P}$
exists and is given by the exponential form (\ref{eq-maxenta}), as
expressed in Proposition~\ref{prop:snob} below.  In the condition, the
notation $a^{T}b$ refers to the dot product between $a$ and $b$.
\begin{description}
\item[Condition 1:] There exists $\pdgnatpar \in \pdgreals^k$ such that
$
Z(\pdgnatpar) = \int_{\pdgoutcome \in \pdgxsamplespace} \exp(-
\pdgnatpar^{T}
T(\pdgoutcome)) d Q(x)
$
is finite and the 
distribution $\tilde{P}$
with density (with respect to $Q$)
\begin{eqnarray}
\label{eq-maxenta}
\tilde{p}(\pdgoutcome) & := & \frac{1}{Z(\pdgnatpar)}e^{-
  \pdgnatpar^{T}
T(\pdgoutcome)} 
\end{eqnarray}
satisfies $E_{\tilde{P}}[T(X)] = \tilde{t}$. 
\end{description}
In our theorems, we shall simply assume that Condition 1 holds. 
A sufficient (by no means necessary!) requirement for Condition 1 is
for example that $Q$ has bounded support; \cite{Csiszar75}
gives a more precise characterization.
We will also assume in our theorems the following natural  condition:
\begin{description}
\item[Condition 2:] 
The `$T$-covariance matrix' $\Sigma$ with 
$\Sigma_{ij} = E_{\tilde{P}}
[ T_{[i]}T_{[j]}] - E_{\tilde{P}}
[ T_{[i]}] E_{\tilde{P}}
[ T_{[j]}]$ is invertible.
\end{description}
$\Sigma$ is guaranteed to exist
by Condition 1 (see any book with a treatment of exponential families, 
for example, \citep{Grunwald07}) and will be singular only
if either $\tilde{t}_{j}$ lies at the boundary of the range of
$T_{[j]}$ for some $j$ or if some of
the $T_{[j]}$ are affine combinations of the others. In the first
case, the constraint $T_{[j]} = \tilde{t}_{j}$ can be replaced by
restricting the sample space to $\{ x \in \pdgxsamplespace \mid T_{[j]}(x)
=\tilde{t}_j \}$ and considering the remaining constraints for the new
sample space. In the second case, we can remove some of the $T_{[i]}$ from
the constraint without
changing the set of distributions satisfying it, 
making $\Sigma$ once again invertible.
\begin{proposition}[\cite{Csiszar75}]\label{prop:snob} 
Assume Condition 1 holds for Constraint (\ref{eq-constraint}). Then \\ 
$\pdgowninf{D( P || Q)}{P : \pdgExp{P}{T(X)} = \tilde{t}} \ $ is
attained by a  $\tilde{P}$ of the form
(\ref{eq-maxenta}). If, additionally, Condition 2 holds, then
Condition 1 holds for 
only one $\pdgnatpar \in \pdgreals^k$ and the infimum is uniquely
attained by the unique $\tilde{P}$ satisfying (\ref{eq-maxenta}).
\end{proposition}
If Condition 1 holds, then $\tilde{t}$ determines both
$\pdgnatpar$ and $\tilde{P}$.  
\section{The Concentration Theorems}
\label{sec:concentration}
\begin{theorem}{\bf (the concentration phenomenon for typical sets, lattice case)}
\label{th-discrete}
Assume we are given a constraint of form (\ref{eq-constraint}) 
such that $T$ is of the lattice form and $h = (h_1, \ldots, h_k)$ is
the span of $T$ and such that conditions 1 and 2 hold. Then  
there exists a sequence $\{c_i\}$ satisfying 
$$\lim_{n \rightarrow \infty} c_n = \frac{\prod_{j=1}^k h_j}{\sqrt{(2 \pi )^k \det
    \Sigma}}
$$ such that
\begin{enumerate}
\item Let $\pdgset{A}_1, \pdgset{A}_2, \ldots$ be an arbitrary
sequence of sets with $\pdgset{A}_i \subset \pdgxsamplespace^i$. For all $n$
with $Q(\pdgssrv_n = \tilde{t}) > 0$, we have:
\begin{equation}
\label{eq:discreteineq}
\tilde{P}(\pdgset{A}_n) \geq   n^{-k/2} c_n Q(\pdgset{A}_n \mid \pdgmyconstraint).
\end{equation}
\end{enumerate}
Hence if $\pdgset{B}_1, \pdgset{B}_2, \ldots$ is a
sequence of sets with $\pdgset{B}_i \subset \pdgxsamplespace^i$ whose probability
tends to $1$ under $\tilde{P}$ in the sense that
$1- \tilde{P}(\pdgset{B}_n) = O(f(n) n^{-k/2})$ for some function $f:
\pdgnaturals \rightarrow \pdgreals$; $f(n) = o(1)$, then $Q(\pdgset{B}_n| \pdgmyconstraint)$
tends to 1 in the sense that $1 - Q(\pdgset{B}_n| \pdgmyconstraint) =
O(f(n)).$
\begin{enumerate} 
\setcounter{enumi}{1}
\item If for all $n$, $\pdgset{A}_n \subseteq \{ x^{(n)} \mid n^{-1}
\pdgstsum T(x_i) = \tilde{t} \}$ then (\ref{eq:discreteineq}) holds with
equality.
\end{enumerate}
\end{theorem}
As discussed in Section~\ref{sec:applications}, 
Theorem~\ref{th-discrete} has applications for data compression.
The relation of the Theorem to Jaynes' original concentration phenomenon
is discussed at the end of the present section.
\begin{proof}
We need the following theorem:
\paragraph{Theorem}{ \bf (`local central limit theorem for lattice
  random variables', \cite{Feller68b}, page 490)}
Let $T = (T_{[1]}, \ldots, T_{[k]})$ be a lattice random vector and $h_1,
\ldots, h_k$ be the corresponding spans 
as in Definition~\ref{def:lattice}; let $E_P [T(X)] = t$ and
suppose that $P$ satisfies Condition 2
with $T$-covariance matrix $\Sigma$. Let $X_1,
X_2, \ldots$ be i.i.d. with common distribution $P$. Let $V$ be a closed and
bounded set in $\pdgreals^k$. Let $v_1, v_2, \ldots$ be a sequence in $V$
such that for all $n$, $P(\pdgstsum (T_i -t) / \sqrt{n} = v_n) > 0$. Then as $n \rightarrow \infty$,
$$
\frac{n^{k/2}}{\prod_{j=1}^k h_j} P \left(\frac{\pdgstsum (T_i-t)}{\sqrt{n}} =
v_n \right) - \pdgnormde(v_n) \rightarrow 0.
$$
Here $\pdgnormde$ is the density of a $k$-dimensional normal distribution
with mean vector $\mu = t$ and covariance matrix
$\Sigma$.

\ 
\\
\noindent
{Feller gives the local central
  limit theorem only
  for 1-dimensional lattice random variables with $E[T] = 0$ and
  $\mbox{var}[T] = 1$; extending the proof to $k$-dimensional random vectors with arbitrary
  means and covariances is, however,
  completely straightforward: see XV.7 (page 494) of
  \citep{Feller68b}.}

\ 
\\
\noindent
The theorem shows that there exists a sequence $d_1, d_2, \ldots$ with
$\lim_{n \rightarrow \infty} d_n =1$ such that, for all $n$ with
$P(\pdgstsum (T_i-t) = \pdgnulvec ) > 0$,
\begin{equation}
\label{eq:discretecrux}
\frac{\frac{n^{k/2}}{\prod_{j=1}^k h_j}P \left(\frac{\pdgstsum
    (T_i-t)}{\sqrt{n}}
= \pdgnulvec \right)}{\pdgnormde(0)} =
\frac{\sqrt{(2 \pi n)^k \det \Sigma}}{\prod_{j=1}^k h_j} P \left(\frac{1}{n}
    \pdgstsum T_i = t \right) = d_n 
\end{equation}
The proof now becomes very simple. First note that $\tilde{P}(\pdgset{A}_n
\mid \pdgmyconstraint) = Q(\pdgset{A}_n \mid \pdgmyconstraint)$ (write out
the definition of conditional probability and realize that $\exp(-
\tilde{\beta}^{T} T(x)) =\exp(-
\tilde{\beta}^{T} \tilde{t}) = \mbox{constant}$ for all $x$
with $T(x) = \tilde{t}$. Use this to show that
\begin{eqnarray}
\label{eq:thesame}
\tilde{P}(\pdgset{A}_n) & \geq & \tilde{P}(\pdgset{A}_n, \pdgmyconstraint) 
=  \tilde{P}(\pdgset{A}_n \mid \pdgmyconstraint) \tilde{P}(\pdgmyconstraint)
\\
& = & Q(\pdgset{A}_n \mid \pdgmyconstraint) \tilde{P}(\pdgmyconstraint) \nonumber. 
\end{eqnarray}
Clearly, with $\tilde{P}$ in the r\^ole of $P$, the local central limit theorem is
applicable to random vector $T$. Then, by (\ref{eq:discretecrux}), 
$\tilde{P}(\pdgmyconstraint) = ({\prod_{j=1}^k h_j})/{\sqrt{(2 \pi n)^k \det
    \Sigma}} d_n$. Defining $c_n := \tilde{P}(\pdgmyconstraint)n^{k/2}$
finishes the proof of item 1.
For item 2, notice that in this case (\ref{eq:thesame}) holds with
equality; the rest of the proof remains unchanged.
\end{proof}
\begin{example}
\label{ex:brandeis}
\rm The `Brandeis dice example' is a toy example frequently used by Jaynes
and others in discussions of the MaxEnt formalism
\citep{Jaynes78}. Let $\pdgxsamplespace = \{1, \ldots, 6 \}$ and $X$ be
the outcome in one throw of some given die. We initially believe
(e.g. for reasons of symmetry) that the distribution of $X$ is
uniform. Then $Q(X = j) = 1/6$ for all $j$ and $E_{Q}[X] =
3.5$. We are then told that the
average number of spots is $E[X]=4.5$ rather than $3.5$. As calculated
by Jaynes, the MaxEnt distribution $\tilde{P}$ given this constraint is
given by 
\begin{equation}
\label{eq:brandeis}
(\tilde{p}(1), \ldots, \tilde{p}(6)) =
(0.05435,0.07877,0.11416,0.16545,0.23977,0.34749).
\end{equation}
By the Chernoff/Hoeffding bound, for every $j \in \pdgxsamplespace$, every $\varepsilon > 0$, 
$\tilde{P}( |n^{-1} \pdgstsum \pdgindicator_{j}(X_i) - \tilde{p}(j) |> \varepsilon)
< 2 \exp(-n c)$ for some constant $c>0$ depending on $\varepsilon$; here
$\pdgindicator_j(X)$ is the indicator function for $X=j$. 
Theorem~\ref{th-discrete} then implies that $Q(
|n^{-1} \pdgstsum \pdgindicator_{j}(X_i) - \tilde{p}(j) |> \varepsilon |
\pdgmyconstraint) = O(\sqrt{n} e^{-nc}) = O(e^{-nc'})$ for some $c' >
0$. In this way we recover Jaynes' original concentration
phenomenon (\ref{eq:snack}): the fraction of sequences satisfying the constraint 
with frequencies close to MaxEnt probabilities $\tilde{p}$ is
overwhelmingly large. 
Suppose now we receive new information about an additional constraint:
$P(X= 4) = P(X=5) = 1/2$.
This can be expressed as a moment constraint by
$E[(I_4(X),I_5(X))^{T}] = (0.5,0.5)^{T}$, where $I_j$ is the indicator
function of the event $X=j$.  We can now
either use
$\tilde{P}$ defined as in (\ref{eq:brandeis}) in the r\^ole of prior
$Q$ and impose the new constraint $E[(I_4(X),I_5(X))^{T}] =
(0.5,0.5)^{T}$, or use uniform
$Q$ and impose the combined constraint
$E[T] = E[(T_{[1]},T_{[2]},T_{[3]})^{T}] =
(4.5,0.5,0.5)^{T}$,
with $T_{[1]} = X, T_{[2]} = I_4(X), T_{[3]} =I_5(X)$. In both
cases we end up with a new  MaxEnt distribution $\tilde{\tilde{p}}(4) =
\tilde{\tilde{p}}(5) = 1/2$.  This distribution, while still consistent
with the original constraint $E[X]= 4.5$, rules out the vast majority
of sequences satisfying it. However, we can apply our concentration
phenomenon again to the new MaxEnt distribution
$\tilde{\tilde{P}}$. Let $\pdgset{I}_{j,j',\varepsilon}$ denote the event that 
$$\left|\frac{1}{n} \pdgstsum \pdgindicator_j(X_i) - 
\frac{\sum_{i=1}^{n-1} \pdgindicator_{j'}(X_i)
  \pdgindicator_j(X_{i+1})}{\sum_{i=1}^{n-1} \pdgindicator_{j'}(X_i)}
\right| > \varepsilon.$$ 
According to $\tilde{\tilde{P}}$, we still have that
$X_1, X_2, \ldots$ are i.i.d. Then by the Chernoff/Hoeffding bound, for each
$\varepsilon > 0$, for $j,j' \in \{4,5\}$, $\tilde{\tilde{P}}(\pdgset{I}_{j,j',\varepsilon})$
is exponentially small.
Theorem~\ref{th-discrete} then implies that
$Q^n(\pdgset{I}_{j,j'\varepsilon} \mid \pdgmyavg = (4.5,0.5,0.5)^{T})$ 
is exponentially small too: for the overwhelming
majority of samples satisfying the combined constraint, the sample
will look just as if it had been generated by an
i.i.d. process, even though $X_1, \ldots, X_n$ are obviously not {\em
  completely\/} independent under $Q^n(\cdot | \pdgmyavg = 
(4.5,0.5,0.5)^{T})$.
\end{example}
There also exists a version of Theorem~\ref{th-discrete}
for continuous-valued random vectors. This is given, along with the
proof, in technical report \citep{Grunwald01b}.

There are a few limitations to Theorem \ref{th-discrete}: (1) we must
require that $\tilde{P}(\pdgset{A}_n)$ goes to $0$ or $1$ as $n \rightarrow \infty$;
(2) the continuous case  needed a
separate statement, which is caused by the more fundamental (3) it
turns out that the
proof technique used cannot be adapted to point-wise
conditioning on $\pdgmyconstraint$ in the continuous case \citep{Grunwald01b}.
Theorem~\ref{th-general} overcomes all these problems. The price we
pay is that, when conditioning on $\pdgmyconstraint$, the sets
$\pdgset{A}_m$ must only refer to $X_1, \ldots, X_m$ where $m$ is such
that $m/n \rightarrow 0$; for example, $m = \lceil n/ \log n \rceil$
will work.
Whenever in the case of continuous-valued $T$ we write $Q(\cdot \mid \pdgmyavg = t)$ or $\tilde{P}(\cdot \mid
\pdgmyavg = t)$ we refer to the continuous version of these quantities.
These are easily shown to exist \citep{Grunwald01b}. Recall that 
(for $m < n$) $Q^{m}(\cdot \mid \pdgmyconstraint)$
refers to the marginal distribution of $X_1, \ldots, X_m$ conditioned
on $\pdgmyconstraint$. It is implicitly understood in the theorem that in
the lattice case, $n$ ranges only over those values for which
$Q(\pdgmyconstraint) > 0$.
\begin{theorem}{ \bf (Main Theorem: the Strong Concentration
    Phenomenon/ Strong Conditional Limit Theorem)}
\label{th-general}
Let $\{m_i \}$ be an increasing  sequence with $m_i \in \pdgnaturals$,
such that $\lim_{n \rightarrow \infty} m_n/n = 0$. Assume we are given a constraint of form (\ref{eq-constraint}) 
such that $T$ is of the regular continuous form  or of the lattice
form 
and suppose that Conditions 1 and 2 are satisfied. Then as $n \rightarrow
\infty$, $Q^{m_n}(\cdot \mid \pdgmyconstraint)$ converges
weakly to $\tilde{P}^{m_n}(\cdot)$.
\end{theorem} 
Discussion of ``weak convergence'' as well as the proof (using the
same key idea, but involving much more work than the proof of
Theorem~\ref{th-discrete}) is in technical report \citep{Grunwald01b}.
%
\paragraph{Related Results}
Theorem~\ref{th-discrete} is related to Jaynes' original concentration
phenomenon, the proof of which is based on Stirling's
approximation of the factorial. 
Another closely related result (also based on
Stirling's approximation) 
is in Example 5.5.8 of
\cite{LiV97}. Both results can be easily extended to prove the following weaker version
of Theorem~\ref{th-discrete}, item 1:
$\tilde{P}(\pdgset{A}_n) \geq n^{-|\pdgxsamplespace|} c_n Q(\pdgset{A}_n |
\pdgmyconstraint)$ where $c_n$ tends to some constant. Note that in this
form, the theorem is void for infinite sample spaces. \cite{Jaynes82}
extends the
original concentration phenomenon in a direction somewhat
different from Theorem~\ref{th-discrete}; it would be interesting to
study the relations. 

Theorem~\ref{th-general} is similar to the original `conditional limit theorems'
(Theorems 1 and 2) of 
\cite{CampenhoutC81}. We note that the preconditions for our theorem
to hold are weaker and the conclusion is stronger than for the
original conditional limit theorems, the main novelty being that
Theorem~\ref{th-general}  supplies us with an
explicit bound on how fast $m$ can grow as $n$ tends to infinity. The
conditional limit theorem was later extended by 
\cite{Csiszar84}. His setting is considerably more general than ours
(e.g. allowing for general convex constraints rather than just moment
constraints), but his results also lack an explicit estimate of the
rate at which $m$ can increase with $n$. \cite{Csiszar84}
and \cite{CoverT91} (where a simplified version of the conditional
limit theorem is proved) both make the connection to large deviation
results, in particular Sanov's theorem. As shown in the latter
reference, weak versions of the conditional limit theorem can be
interpreted as immediate consequences of Sanov's theorem.
\section{Consequences for Data Compression Games}
\label{sec:applications}
For simplicity we  restrict ourselves in this section to countable sample spaces
$\pdgxsamplespace$  and we identify
probability mass functions with probability distributions.
Below we make frequent use of
coding-theoretic concepts which we first briefly review.
\subsection{Theorem 1 and Data Compression}
Recall that by the Kraft Inequality \citep{CoverT91}, for every prefix
code with lengths $L$ over symbols from a countable alphabet $\pdgxsamplespace^n$,
there exists a (possibly sub-additive) 
probability mass function $p$ over $\pdgxsamplespace^n$ such
that for all $x^{(n)} \in \pdgxsamplespace^{n}$, $L(x^{(n)}) = - \log
p(x^{(n)})$. We will call this $p$ the `probability (mass) function
corresponding to $L$'. Similarly, for every probability mass function $p$
over $\pdgxsamplespace^n$ there exists a (prefix) code with lengths
$L(x^{(n)}) = \lceil - \log
p(x^{(n)}) \rceil$. Neglecting the round-off error, we will simply say
that for every $p$, there exists a code with  lengths 
$L(x^{(n)}) = - \log p(x^{(n)})$. We call the code with these lengths 
`the code corresponding to $p$'. By the
information inequality \citep{CoverT91}, this is also the most efficient code to use if
data $X^{(n)}$ were actually distributed according to $p$. 

We can now see that 
Theorem~\ref{th-discrete}, item 2, has important implications
for coding. Consider the following
special
case of Theorem~\ref{th-discrete}, which obtains by taking $\pdgset{A}_n = \{ x^{(n)}\}$ and
logarithms:
\begin{corollary}{\bf (the concentration phenomenon, coding-theoretic formulation)}
\label{corr:discrete}
Assume we are given a constraint of form (\ref{eq-constraint}) 
such that $T$ is of the lattice form and $h = (h_1, \ldots, h_k)$ is
the span of $T$ and such that conditions 1 and 2 hold. For all $n$,
all $x^{(n)}$ with $n^{-1} \pdgstsum T(x_i) = \tilde{t}$, we have 
\begin{eqnarray}
\label{eq:k2logn}
&- \log \tilde{p}(x^{(n)}) = - \log q(x^{(n)} \mid
\frac{1}{n} \pdgstsum T(X_i) = \tilde{t})  + & \nonumber \\ & + \frac{k}{2} \log {2
  \pi n} + \log \sqrt{ \det \Sigma} -
\sum_{j=1}^k \log h_j + o(1) = & \nonumber \\ 
& - \log q(x^{(n)} \mid
\frac{1}{n} \pdgstsum T(X_i) = \tilde{t}) + \frac{k}{2} \log n + O(1). &
\end{eqnarray}
\end{corollary}
In words, this means the following: let $x^{(n)}$ be a sample
distributed according to $Q$, Suppose we are given the information that $n^{-1} \pdgstsum T(x_i) = \tilde{t}$. Then, by the information inequality, the most efficient
code to encode $x^{(n)}$ is the one based on $q(\cdot | \pdgmyconstraint)$
with lengths $- \log q(x^{(n)}\mid \pdgmyconstraint)$. Yet if 
we encode $x^{(n)}$ using the code with lengths $- \log \tilde{p}(\cdot)$
(which would be the most efficient had $x^{(n)}$ been generated by
$\tilde{p}$) then the number of extra
bits we need is only of the order $(k/2) \log n$. That means, for
example, that the
number of additional bits we need {\em per outcome\/} goes to $0$ as
$n$ increases.
\cite{Grunwald01a} used Corollary~\ref{corr:discrete} to
establish a formal connection between the concentration phenomenon and
{\em universal coding}, a central concept of information theory; this
is worked out in more detail by  \cite{Grunwald07}, Chapter 10,
Section 2.2.  In the present paper, we focus on the game-theoretic
consequences of Corollary~\ref{corr:discrete}. 
\subsection{Empirical Constraints and \mbox{Game~Theory}}
\label{sec:game}
Recall we assume countable $\pdgxsamplespace$. The $\sigma$-algebra of
such $\pdgxsamplespace$ is tacitly taken to be the power set of
$\pdgxsamplespace$. The $\sigma$-algebra thus being implicitly
understood, we can define $\pdgprobset(\pdgxsamplespace)$ to be the set of
all probability distributions over $\pdgxsamplespace$. For a product
${\cal X}^{\infty} = \times_{i \in \pdgnaturals} {\cal X}$ of a countable
sample space ${\cal X}$, we define $\pdgprobset(\pdgxsamplespace^{\infty})$
to be the set of all distributions over the product space with the
associated product $\sigma$-algebra.

\cite{Topsoe79} and \cite{GrunwaldD04} provided characterizations of Maximum Entropy
distributions quite different from the present one.
It was shown that, under regularity conditions,
\begin{equation}
  \label{eq:minimax}
\pdgmaxent  \ = \ \sup_{p^*: E_{p^*} [T] = \tilde{t}} \ \ \inf_{p} \ \
  E_{p^*}\left[ - \log \frac{p(X)}{q(X)}\right] = \ 
\inf_{p} \ \ \sup_{p^*: E_{p^*} [T] = \tilde{t}} \ \ E_{p^*} \left[ - \log \frac{p(X)}{q(X)}\right]
\end{equation}
where both $p$ and $p^*$ are understood to be members of
$\pdgprobset(\pdgxsamplespace)$ and $\pdgmaxent$ is defined as in
(\ref{eq:maxentdef}).  By this result, the MaxEnt setting can be
thought of as a game between Nature, who can choose any $p^*$
satisfying the constraint, and Statistician, who only knows that
Nature will choose a $p^*$ satisfying the constraint. Statistician
wants to minimize his worst-case expected codelength (relative to
$q$), where the worst-case is over all choices for Nature. In such
game-theoretic contexts, the codelength is usually called
``logarithmic score'' or ``logarithmic loss'' \citep{GrunwaldD04}.

It turns out that the minimax strategy for Statistician in
(\ref{eq:minimax}) is
given by $\tilde{p}$. That is,
\begin{equation}
  \label{eq:arginf}
  \tilde{p} = \arg \inf_p \ \sup_{p^*: E_{p^*}[T] = \tilde{t}} \ \
  E_{p^*} 
\left[ -
  \log \frac{p(x)}{q(x)} \right].
\end{equation}
Thus, $\tilde{p}$ is both the optimal strategy for Nature and for
Statistician.  This gives a decision-theoretic justification of using
MaxEnt probabilities which seems quite different from our
concentration phenomenon. Or is it? Realizing that in practical
situations we deal with empirical constraints of form (\ref{eq-emp})
rather than (\ref{eq-constraint}) we may wonder what distribution
$\hat{p}$ is minimax in the empirical version of problem
(\ref{eq:arginf}). In this version Nature gets to choose an individual
sequence rather than a distribution. To our knowledge, we are
  the first to analyze this `empirical' game. To make it more precise,
let
\begin{equation}
\label{eq:constraintset}
{\cal C}_n = \left\{ 
x^{(n)} \in \pdgxsamplespace^n \mid  n^{-1} \pdgstsum
T(x_i) = \tilde{t} \right\}. 
\end{equation}
Then, for $n$ with ${\cal C}_n \neq \emptyset$, 
$\hat{p}_n$ (if it exists) is defined by
\begin{equation}
  \label{eq:empinfsup}
  \hat{p}_n :=
\arg \inf_{p \in \pdgprobset(\pdgxsamplespace^{n})}
\ \ \sup_{x^{(n)} \in {\cal C}_n} \ - 
\log \frac{p(x_1,\ldots, x_n)}{q(x_1,\ldots, x_n)}
=
\arg \sup_{p \in \pdgprobset(\pdgxsamplespace^{n})}
\ \ \inf_{x^{(n)} \in {\cal C}_n}
\ \frac{p(x^{(n)})}{q(x^{(n)})}
\end{equation}
$\hat{p}_n$ can be interpreted in two ways: (1) it is the distribution that
assigns `maximum probability' (relative to $q$) to all sequences
satisfying the constraint; (2) as $- \log
(\hat{p}(x^{(n)})/q(x^{(n)})) = \pdgstsum (- \log \hat{p}(x_i| x_1,
\ldots, x_{i-1}) + \log  q(x_i| x_1,
\ldots, x_{i-1}))$, it is also the $p$ that minimizes cumulative
worst-case
logarithmic loss relative to $q$ when used for sequentially predicting $x_1,
\ldots, x_n$.

One immediately verifies that $\hat{p}_n = q^{n}(\cdot \mid
\pdgmyconstraint)$: the solution to the empirical minimax
problem is just the conditioned prior, which we know
by Theorems~\ref{th-discrete} and \ref{th-general} is in some sense
very 
close to
$\tilde{p}$. However, for no single $n$, is  $\tilde{p}$ exactly equal to $q^{n}(\cdot \mid
\pdgmyconstraint)$. Indeed, $q^{n}(\cdot \mid
\pdgmyconstraint)$ assigns zero  probability to any sequence of length $n$
not satisfying the constraint. This means that using $q$ in prediction tasks against the
logarithmic loss will be problematic if the constraint only
holds approximately 
and/or if $n$ is unknown in advance to the Statistician. In the latter case,
it is impossible to use $q(\cdot \mid
\pdgmyconstraint)$ for prediction without modification. For suppose
that the statistician guesses that the sample will have length $n_1$
for some $n_1$ with ${\cal C}_{n_1} \neq \emptyset$. There exist sequences $x^{(n_2)} = x_1, \ldots, x_{n_1}, \ldots, x_{n_2}$ of length $n_2 > n_1$
satisfying the constraint such that $x^{(n_1)}$ does not satisfy the
constraint, and therefore $q(x^{(n_2)}| x^{(n_1)} \in
{\cal C}_{n_1}) = 0$, so $q(\cdot \mid  x^{(n_1)} \in
{\cal C}_{n_1}) = 0$ cannot be used for prediction if the actual
sequence length turns out to exceed $n_1$. We may guess that in this case ($n$
not known in advance), the MaxEnt distribution
$\tilde{p}$, rather than $q(\cdot | \pdgmyconstraint)$ is actually a better
distribution to use for prediction. The following theorem shows that
in some sense, 
this is indeed so:
\begin{theorem}
\label{th:infsup}
Let $\pdgxsamplespace$ be a countable sample space. Assume we are given a constraint of form (\ref{eq-constraint}) 
such that $T$ is of the lattice form, and such
that Conditions 1 and 2 are satisfied. Let ${\cal C}_n$ be
as in (\ref{eq:constraintset}). Then the infimum in
\begin{equation}
\label{eq:infsupdata}
\inf_{p \in \pdgprobset(\pdgxsamplespace^{\infty})} \ 
\sup_{\{n\; :\; {\cal C}_n \neq
\emptyset\}} \  \sup_{x^{(n)} \in {\cal C}_n} \ - 
\frac{1}{n} \log \frac{p(x_1,\ldots, x_n)}{q(x_1,\ldots, x_n)}
\end{equation}
is achieved by 
the Maximum Entropy distribution
$\tilde{p}$, and is equal to $\pdgmaxent$. 
\end{theorem}
\begin{proof}
Let
${\cal C} = \cup_{i=1}^{\infty} {\cal C}_i$. 
We need to show that for all $n$, for all $x^{(n)} \in {\cal C}$,
\begin{equation}
  \label{eq:maxentisinfsup}
\pdgmaxent = -\frac{1}{n} \log \frac{\tilde{p}(x^{(n)})}{q(x^{(n)})} 
= \inf_{p \in \pdgprobset(\pdgxsamplespace^{\infty})} \sup_{\{n\; :\; {\cal C}_n \neq
\emptyset\}}
\sup_{x^{(n)} \in {\cal C}_n} - 
\frac{1}{n} \log \frac{p(x^{(n)})}{q(x^{(n)})}
\end{equation}
Equation~(\ref{eq:maxentisinfsup}) implies that $\tilde{p}$ reaches the
$\inf$ in (\ref{eq:infsupdata}) and that the $\inf$ is equal to
$\pdgmaxent$. 
The leftmost equality in (\ref{eq:maxentisinfsup}) is a standard
result about exponential families of form (\ref{eq-maxenta}); see for example,
\citep{Grunwald07}.
To prove the rightmost equality in (\ref{eq:maxentisinfsup}), 
let $x^{(n)} \in {\cal C}_n$. 
Consider the conditional distribution $q(\cdot \mid x^{(n)} \in {\cal C}_n)$. Note that, for every distribution $p_0$ over
$\pdgxsamplespace^n$, $p_0(x^{(n)}) \leq q(x^{(n)}| x^{(n)} \in {\cal C}_n)$ for at least one
$x^{(n)} \in {\cal C}_n$. By Theorem~\ref{th-discrete} (or
rather Corollary~\ref{corr:discrete}), 
for this $x^{(n)}$ we have
$$
- \frac{1}{n} \log 
\frac{p_0(x^{(n)})}{q(x^{(n)})} 
\geq - \frac{1}{n} \log \frac{\tilde{p}(x^{(n)})}{q(x^{(n)})} 
-
\frac{k}{2n} \log n - O(\frac{1}{n}),
$$
and we see that for every distribution $p_0$ over $\pdgxsamplespace^{\infty}$, 
$$\sup_{\{n\; :\; {\cal C}_n \neq
\emptyset\}}  \ \ 
\sup_{x^{(n)} \in {\cal C}_n} - 
\frac{1}{n} \log\frac{p_0(x^{(n)})}{q(x^{(n)})} 
\geq \sup_{\{n\; :\; {\cal C}_n \neq
\emptyset\}} 
\ \  \sup_{x^{(n)} \in {\cal C}_n} - 
\frac{1}{n} \log\frac{\tilde{p}(x^{(n)})}{q(x^{(n)})},$$
which shows the rightmost equality in
(\ref{eq:maxentisinfsup}).
\end{proof}
Theorem~\ref{th:infsup} shows that, among all distributions on ${\cal
  X}^\infty$, the minimax codelength {\em per outcome\/} for sequences
satisfying the constraints is achieved by the maximum entropy
$\tilde{p}$. We may now ask whether it is also achieved by any
different distribution $p'$, and if so, whether that distribution may
even be ``better'' in the sense that it achieves strictly smaller
codelengths on all sequences of all lengths that satisfy the
constraints. Surprisingly, the answer depends on the number of
constraints $k$: for $k > 2$, there exists such a $p'$. For $k \leq
2$, there does not:
\begin{theorem}
\label{thm:A}
Assume we are given a constraint such that Condition 1 and
2 both hold, ${\cal X}$ is finite, $T$ is of the lattice form, and $T = (T_{[1]}, \ldots, T_{[k]})$ for some $k >
2$. Then (a), there exists a distribution $p'$ and a constant $c' > 0$,
such that, for all large enough $n$
with ${\cal C}_n \neq \emptyset$, for all 
$x^{(n)} \in {\cal C}_{n}$,
$$\log \frac{p'(x_1, \ldots,
  x_n)}{\tilde{p}(x_1, \ldots, x_n)} >  c \log n. 
$$
Moreover, (b), there exists a distribution $p''$ and a constant $c'' >
0$ such that for {\em all\/} $n$ (and not just all large $n$) 
with ${\cal C}_n \neq \emptyset$, for all 
$x^{(n)} \in {\cal C}_{n}$,
$$\log \frac{p''(x_1, \ldots,
  x_n)}{\tilde{p}(x_1, \ldots, x_n)} > c''. 
$$
\end{theorem}
\begin{theorem}
\label{thm:B}
Assume we are given a constraint such that Condition 1 and
2 both hold, ${\cal X}$ is finite, $T$ is of the lattice form, and $T = (T_{[1]}, \ldots, T_{[k]})$ for some $k \leq
2$. Then there exists {\em no\/} distribution $p'$,
such that, for all large  $n$
with ${\cal C}_n \neq \emptyset$, for all 
$x^{(n)} \in {\cal C}_{n}$,
$$\log \frac{p'(x_1, \ldots,
  x_n)}{\tilde{p}(x_1, \ldots, x_n)} > 0.
$$
\end{theorem}
The upshot of these theorems is that, if it is known that the sample
satisfies the constraint, but the sample size is not known, then, if
$k \geq 3$, there exist distributions which are guaranteed to compress
the data more than $\tilde{p}$, so that the game-theoretic
justification for predicting/coding with $\tilde{p}$ is, to some
extent, challenged. 
The proofs of both theorems make use of the following lemma, which we
state and prove first:
\begin{lemma}
\label{lem:bem}
Under the conditions of Theorem~\ref{thm:A} and~\ref{thm:B}, suppose there exists a distribution $p'$
such that, for all large enough $n$ with ${\cal C}_n \neq \emptyset$, for all 
$x^{(n)} \in {\cal C}_{n}$,
\begin{equation}
\label{eq:klef}
\log \frac{p'(x_1, \ldots,
  x_n)}{\tilde{p}(x_1, \ldots, x_n)} > 0. 
\end{equation}
Then there also exists a
distribution $p''$ and a fixed $c'' > 0$ such that for {\em all\/} $n$ (and not just for
all large $n$)
with ${\cal C}_n \neq \emptyset$, for all 
$x^{(n)} \in {\cal C}_{n}$,
$\log \frac{p''(x_1, \ldots,
  x_n)}{\tilde{p}(x_1, \ldots, x_n)} > c''. $
\end{lemma}
\begin{proof}
  Let $n^*$ be the smallest $n$ such that for all $x^{(n)} \in {\cal
  C}_n$,  condition (\ref{eq:klef})
  holds. Let $n_1$ be the smallest $n$ such that ${\cal C}_n$ is
  nonempty. Note that $n^* \geq  n_1$.
  Let $a = \lceil n^*/n_1\rceil$, i.e. $n^*/n_1$ rounded up to the
  nearest integer. Then $n_1 \cdot a \geq n^*$, so that, with $n_2 :=
  n_1(a-1)$, we have $0 \leq  n_2 < n^*$. We only consider the case $0
  < n_2$; the case $n_2 = 0$ is completely analogous, but easier. So
  assume $n_2 > 0$. Then  ${\cal C}_{n_2}$ is nonempty (to
  see this, take any $x^{(n_1)} \in {\cal C}_{n_1}$ and note that the
  sequence consisting of $(a-1)$ repetitions of $x^{(n_1)}$ must
  satisfy the constraint and therefore be in ${\cal C}_{n_2}$).  We
  may assume that
\begin{equation}
\label{eq:snef}
\inf_{x^{(n_2)} \in {\cal C}_{n_2}}
 \frac{p'(x^{(n_2)})}{\tilde{p}(x^{(n_2)})} \leq 1,
\end{equation}
 (otherwise it would follow that $n_2$
  rather than $n^*$ is the smallest $n$ for which condition
  (\ref{eq:klef}) holds, and we would have a contradiction). 
 Now, let $y^{(n_2)} \in {\cal C}_{n_2}$ be any sequence that achieves
the infimum in (\ref{eq:snef}). Since by our condition, for all $x^{(a \cdot n_1)} \in {\cal
  C}_{a \cdot n_1}$, $p'(x^{(a \cdot n_1)})/\tilde{p}( x^{(a \cdot
  n_1)}) > 1$, and also $p'(x^{(a n_1)}) =  p'(x_{n_2 +1},
  \ldots, x_{a n_1} \mid x^{(n_2)}) \cdot p'(x^{(n_2)})$, and also
  ${\cal C}_{n_1}$ is finite, it follows
  that 
\begin{equation}
\label{eq:wiske}
\inf_{z^{(n_1)} \in {\cal C}_{n_1}}
\frac{P'(X_{n_2 +1} = z_1, \ldots, X_{a n_1} = z_{n_1} \mid X^{(n_2)} =
y^{(n_2)})}{\tilde{p}(z^{(n_1)})} > 1.
\end{equation}
We may now define a probability mass function $p^\circ$ on ${\cal
  X}^{n_1}$ such that, for $z^{(n_1)} \in {\cal C}_{n_1}$,
$p^\circ(z^{(n_1)})$ is slightly smaller than $P'(X_{n_2 +1} = z_1,
\ldots, X_{a n_1} = z_{n_1} \mid X^{(n_2)} = y^{(n_2)})$, whereas for
$z^{(n_1)} \in {\cal X}^{n_1} \setminus {\cal C}_{n_1}$,
$p^\circ(z^{(n_1)})$ is slightly larger than $P'(X_{n_2 +1} = z_1,
\ldots, X_{a n_1} = z_{n_1} \mid X^{(n_2)} = y^{(n_2)})$, where we
increase and decrease the probability in such a way that the total
probability on ${\cal X}^{n_1}$ remains 1. Since ${\cal X}^{n_1}$ is
finite, using (\ref{eq:wiske}), we can do this in such a way that for some $\varepsilon > 0$, for
all $z^{(n_1)} \in {\cal C}_{n_1}$,
\begin{equation}
\label{eq:vlef}
\frac{p^\circ(z^{n_1})}{\tilde{p}(z^{(n_1)})} \geq 1 + \varepsilon,
\end{equation}
whereas for all $z^{(n_1)}  \in {\cal X}^{n_1} \setminus {\cal
  C}_{n_1}$,
\begin{equation}
\label{eq:vlefb}
\frac{p^\circ(z^{n_1})}{P'(X_{n_2 +1} = z_1, \ldots, X_{a n_1} = z_{n_1} \mid X^{(n_2)} =
y^{(n_2)})} \geq 1 + \varepsilon,
\end{equation}
We now extend $p^\circ$ to a probability mass function on ${\cal
  X}^\infty$ by defining, for all $z^{(n_1)} \in {\cal X}^{n_1}$, for any $m \geq 1, y^{(m)} \in {\cal X}^m$,
$p^\circ(z_1, \ldots, z_{n_1}, y_1, \ldots, y_m) := p^\circ(z^{(n_1)})
  P'(X_{n_1 + 1} = y_1, \ldots, X_{n_1 + m} = y_m \mid X^{n_1} = z^{(n_1)})$.
Now, for any $n \geq 1$, $x^{(n)} \in {\cal X}^n$, let $m$ be the
number of distinct initial segments $x_1, \ldots, x_{n'}$ of $x^{(n)}$
  with $n' < n$
that satisfy the constraint, i.e. $x^{(n')} \in {\cal C}_{n'}$. Notice
  that we may have $m=0$. We set $s_0 = 0$, $s_{m+1} = n$, and, for $j \in \{1,
\ldots, m \}$, $s_j$ is set such that $x^{(s_j)} \in {\cal C}_{s_j}$ and $s_1 < s_2 <
\ldots < s_m < n$. We define
$$
p''(x^{(n)}) := \prod_{j=0}^{m} p^\circ(x_{s_j+1}, \ldots, x_{s_{j+1}} ). 
$$ 
One easily verifies by induction on $n$ that $p''$ defines a probability mass function on
${\cal X}^\infty$, and, using (\ref{eq:snef}) and (\ref{eq:vlefb}),
that for all $n$ with ${\cal C}_n \neq \emptyset$, all $x^n \in {\cal C}_n$,
$p''(x^n)/\tilde{p}(x^n) \geq 1 + \varepsilon$. The result follows.
\end{proof}
\begin{proof}{\bf (of Theorem~\ref{thm:A})}
Let $n_1 < n_2 < \ldots$ be the sequence of all $n$ such that ${\cal
  C}_n \neq \emptyset$. Define, for $j= 1, 2, \ldots$, $q_j := q(\cdot
\mid \pdgaverage{T}{(n_j)}=t)$. Let $\pi$ be any distribution on
the natural numbers such that, for all $j \in \{1, 2, \ldots \}$,
$$
- \log \pi(j) = \log j + O (\log \log j).
$$
For example, we may take Rissanen's {\em universal prior for the
  integers\/} \citep{Rissanen89}, $\pi(j) \propto 1 / j
(\log j)^2$.
Now set, for all $n$, $x^n \in {\cal
X}^n$,  $p'(x^n) := \sum_{j = 1, 2, \ldots} \pi(j) q_j(x^n)$.
Then $p'$ uniquely induces a distribution $P$ on ${\cal X}^\infty$ that is actually a
mixture of {\em all\/} conditional distributions of ${\cal X}^\infty$ given
that the constraint holds at some sample size for which it can hold
at all. For each $j \in \{1, 2 \ldots \}$, each $x^{(n_j)}
\in {\cal C}_{n_j}$, we have 
\begin{eqnarray}
- \log p'(x^{(n)}) & = & - \log \sum_{i} \pi_i q_i(x^{(n)}) \nonumber
\\ & \leq & - \log \pi(j) - \log q(x^{(n)}
\mid \pdgaverage{T}{(n_j)}=t) \nonumber 
\\ & \leq & \log j + O(\log \log j) - \frac{k}{2}
\log n - \log \tilde{p}(x^{(n)}), \nonumber
\end{eqnarray}
where the final inequality follows by Corollary~\ref{corr:discrete}.
Since $j \leq n$, Part (a) of the theorem  follows. Part (b) is now an
immediate consequence of Lemma~\ref{lem:bem}.
\end{proof}
\begin{proof}{\bf (of Theorem~\ref{thm:B})}
Assume, by means of contradiction, that a $p'$ as mentioned in the
theorem does exist. Then by Lemma~\ref{lem:bem}, there also exists a
$p''$, such that for all $n$
with ${\cal C}_n \neq \emptyset$, for all 
$x^{(n)} \in {\cal C}_{n}$,
$\log \frac{p''(x_1, \ldots,
  x_n)}{\tilde{p}(x_1, \ldots, x_n)} > c''$ for some $c'' > 0$.
For $0 < \alpha < 1$, define a probability distribution on ${\cal X}^\infty$, 
in terms of its mass function $p_{\alpha}$, by $p_{\alpha}(x^n) :=
\alpha p''(x^n) + (1- \alpha) \tilde{p}(x^n)$.  Note that
\begin{equation}
\label{eq:laat}
p_{\alpha}(x^n) \geq \max \{ \alpha p''(x^n), (1- \alpha) \tilde{p}(x^n) \}.
\end{equation}

Now, for any $n \geq 1$, $x^{(n)} \in {\cal X}^n$, let $m$ be the
number of distinct initial segments $x_1, \ldots, x_{n'}$ of $x^{(n)}$
with $n' < n$
that satisfy the constraint, i.e. $x^{(n')} \in {\cal C}_{n'}$. Notice
  that we may have $m=0$. We set $s_0 = 0$, $s_{m+1} = n$, and, for $j \in \{1,
\ldots, m \}$, $s_j$ is set such that $x^{(s_j)} \in {\cal C}_{s_j}$ and $s_1 < s_2 <
\ldots < s_m < n$. We define
$$
p^\circ(x^{(n)}) := \prod_{j=0}^{m} p_{\alpha}(x_{s_j+1}, \ldots, x_{s_{j+1}}). 
$$
One easily verifies by induction on $n$ that (a)  $p^{\circ}$ is the mass
function of some probability distribution $P^{\circ}$ on ${\cal
  X}^\infty$, and, using (\ref{eq:laat}), that (b) for all $n$, all $x^n \in {\cal X}^n$,
\begin{eqnarray}
\label{eq:plas}
p^\circ(x^{(n)}) & \geq  &  \left( \prod_{j=0}^{m-1} \alpha p''(x_{s_j+1}, \ldots,
x_{s_{j+1}}) \right) (1- \alpha) \tilde{p}(x_{s_m +1, \ldots, x_n})
\nonumber \\ &
\geq &  \alpha^m e^{m c''} \tilde{p}(x_1, \ldots, x_{s_m}) (1- \alpha)
\tilde{p}(x_{s_m +1, \ldots, x_n}) \nonumber \\ & = & 
\alpha^m (1- \alpha) 2^{m c''} \tilde{p}(x^{(n)}),
\end{eqnarray}
where $c'' $ is as in Lemma~\ref{lem:bem}. 
Now suppose that $X_1, X_2, \ldots$ are i.i.d. $\sim \tilde{P}$, i.e. data
are sampled from the MaxEnt distribution $\tilde{P}$. We may view
$U_n := n^{-1} \sum_{i=1}^n T(X_i)  - \tilde{t}$
as specifying a Markov chain, where the state at time $n$ is given by
the value of $U_n$, and the transition probabilities are given by 
$\tilde{P}(U_{n+1} = \cdot \mid U_{n} = u )$, for each
realizable value of $u$, and the starting state is $U_0 := 0$. By the local central limit
theorem (Section~\ref{sec:concentration}), the probability of being in
state ``0'' at time $n$ is of order $1 / \sqrt{n}$ (if $k=1$) or
$1 /n$ (if $k=2$). In both cases, this probability is summable, so it
follows by basic Markov chain theory \citep{Feller68b} that state ``0''
is
recurrent and with probability 1, $U_n = 0$ will
  hold for infinitely many $n$. But, for $n > 0$,  $U_n = 0$ is equivalent to
  $x^{(n)} \in {\cal C}_n$, i.e. the constraint holds. It follows that
    the constraint will hold infinitely often, almost surely under
    $\tilde{P}$. 
Yet, if we decide to encode sequence $x_1, \ldots, x_n$ with the code
corresponding to $p^\circ$ rather than $\tilde{p}$, then 
by (\ref{eq:plas}), if we select a value of $\alpha < 1$ such that $\alpha
2^{c''} > 1$, then we will $\tilde{P}$-almost surely compress the data
significantly better than if we use the code with lengths $- \log \tilde{p}$
itself.  More precisely, with $\tilde{P}$-probability 1,
$$
- \log \frac{\tilde{p}(X^n)}{p^\circ(X^n)} \rightarrow \infty.
$$
But this contradicts the no-hypercompression inequality
\citep{Grunwald07} (also known as the ``competitive optimality of the
Shannon-Fano code, '' \citep{CoverT91}), an easy consequence of Markov's inequality 
which states that for all $K> 0$, and any two distributions $P$ and
$Q$ with mass functions $p$ and $q$, for all $n$,
$P(- \log p(X^n) \geq - \log q(X^n) + K) \leq 2^{-K}$. The theorem is proved.
\end{proof}
\bibliographystyle{plainnat}
\bibliography{master,peter,MDL}
\end{document}